\documentclass{article}

\usepackage{apalike}
\usepackage{latexsym}
\usepackage{graphicx}
\usepackage{mathptmx}
\usepackage{lipsum}
\usepackage{amsmath}
\usepackage{algorithm}
\usepackage{algpseudocode}

\usepackage{amsmath}
\usepackage{amsfonts}
\usepackage{amssymb}
\usepackage{amsbsy}
\usepackage{amsthm}

\usepackage[pdftex,colorlinks=true,urlcolor=blue,citecolor=black,anchorcolor=black,linkcolor=black,bookmarks=false]{hyperref}

\usepackage{hyphenat}
\newtheoremstyle{scsthe}
{8pt}
{8pt}
{\it}
{}
{\bf}
{.}
{.5em}
{}

\theoremstyle{scsthe}

\begin{document}

\title{REAL-TIME ADAPTIVE ABSTRACTION AND APPROXIMATION USING VALIDITY FRAMES - AN EXPERIENCE REPORT}

\author{
\\
Raheleh Biglari \\
Joachim Denil \\[12pt]
Department of Electronics and ICT \\
Faculty of Applied Engineering \\
Cosys-lab, University of Antwerp \\
Flanders Make@Uantwerpen\\
Groenenborgerlaan 171,
Antwerp, Belgium \\
\{raheleh.biglari,\\
joachim.denil\}@uantwerpen.be\\
}

\date{} 
\maketitle

\section*{ABSTRACT}
Designing a Cyber-Physical System (CPS), including modeling the control components and services, is a challenging task. Using models and simulations during run-time is crucial for successfully implementing advanced control and prediction components. The complexity of designing an effective CPS system increases due to real-time constraints. Generating accurate predictions and making decisions using detailed models in various contexts is 
computationally demanding and complex to manage within the available computational resources. Employing approximated models and switching to the most suited model adaptively at run-time is an effective technique. But an approximated model is most probable not valid in all the different contexts the system will be in. This experience report uses the Validity Frame concept to enable 
adaptation at run-time. In each environment, some influencing factors are outside the model's control, but these properties influence the model's behavior. By defining Validity Frames, based on specific contexts and related models, we present a possible perspective to address the issue of selecting the more appropriate model in various contexts. Furthermore, we discuss the insights and lessons obtained and determine future challenges.

\textbf{Keywords:} Cyber-Physical Systems (CPS), Model-Based Systems Engineering (MBSE), Validity Frames, Adaptive Approximation, Real-time.

\section{INTRODUCTION} \label{sec:intro}
In the design and development of software-intensive systems, particularly in the avionics and automotive industries, engineers are faced with the challenge of dealing with highly complex systems composed of various interrelated and deeply integrated components \cite{lee2008cyber}. These systems, known as Cyber-Physical Systems (CPS) typically operate in harsh conditions with rapidly fluctuating environmental conditions. In these dynamic environments, the ability to predict the behavior and performance of the system is often not guaranteed\cite{palumbo2017challenging}.\\ 
To address the challenges presented by a changing environment, adaptivity is integrated into the design of CPSs to allow the system to respond to environmental changes. This adaptation must typically be performed in a decentralized manner due to the autonomous nature of many real-time CPSs, which increases the overall complexity of designing these adaptive systems \cite{kit2015architecture}.\\ 
Cyber-Physical Systems, which integrate embedded control, mechanics, and networking, are typically developed using a model-based approach. This approach involves utilizing one or more physics-based models for designing and simulating the system. Since a CPS is a complicated system, its model also tends to be complicated. Additionally, these models are employed for control of the system at run-time. Complicated models have a heavy computational load on the embedded devices that typically implement the control on the system. As such, implementing these computationally demanding models is challenging, possibly resulting in missed deadlines and increased costs. 
One way to balance improved performance and cost reduction in CPS is to use adaptive abstraction and approximation techniques \cite{9004843,8904655,biglari2022towards}. Adaptive approximation specifically switches between multiple prediction models to achieve improved computational efficiency while maintaining an acceptable level of control performance.
The adaptive approximation technique enables the utilization of a simpler, less detailed model with a lower computational cost.\\

In this experience report, we applied the idea of utilizing the Validity Frame to ease the development of real-time Cyber-Physical Systems while using the run-time adaptive abstraction and approximation technique.
We use the properties of interest and environment properties which are known as influencing factors within a Validity Frame. Moreover, we exploit the relations between different Validity Frames to choose the most appropriate
model from the library of models.\\

The structure of this paper is as follows: section \ref{sec: related} provides the related work. Next, section \ref{sec: report} introduces our case study and details the adaptive approximation method applied to the system under study. Afterward, section \ref{sec:discussion} discusses the challenges encountered, lessons gained, and insights of this research.
Finally, section \ref{sec:conclusion} concludes with a summary and future research based on the insights gained.

\section{RELATED WORK} \label{sec: related}
This section describes related work for both adaptive abstraction and approximation and validity frames.

\subsection{Adaptive Abstraction and  Approximation}
Due to the complexity of CPSs, CPS models also possess similar complexity. Utilizing models with heavy computational requirements is challenging as it can cause missing deadlines.

To avoid computationally expensive models, there is a possibility of eliminating the reasoning on a set of properties within a model. This is known as a more abstract model. As the model removes certain properties, it is typically also less computationally expensive to simulate. 
An alternative way to exclude factors from a model is the approximation of resolution which means removing detail, while retaining the set of properties one can reason on with the model. As such a more approximate model uses a simpler model to approximate the results of a more complicated model, assuming the difference between the two is insignificant for the results of the simulation \cite{frantz1996model}.

It is possible to have multiple abstracted and approximated models. To switch between these models at run-time, we benefit from the idea of adaptation by \cite{MATHIEU201851,8904655,9004843} that dynamically switches between abstractions during simulation or execution or simulation.

Therefore, the adaptive approximation technique allows for the use of a simpler, less detailed model with a lower computational cost. Allowing adaptivity at run-time is one technique for resolving
the conflict between better performance and reduced cost in CPS.

\begin{figure}
    \centering
    \includegraphics[width = 0.5\columnwidth]{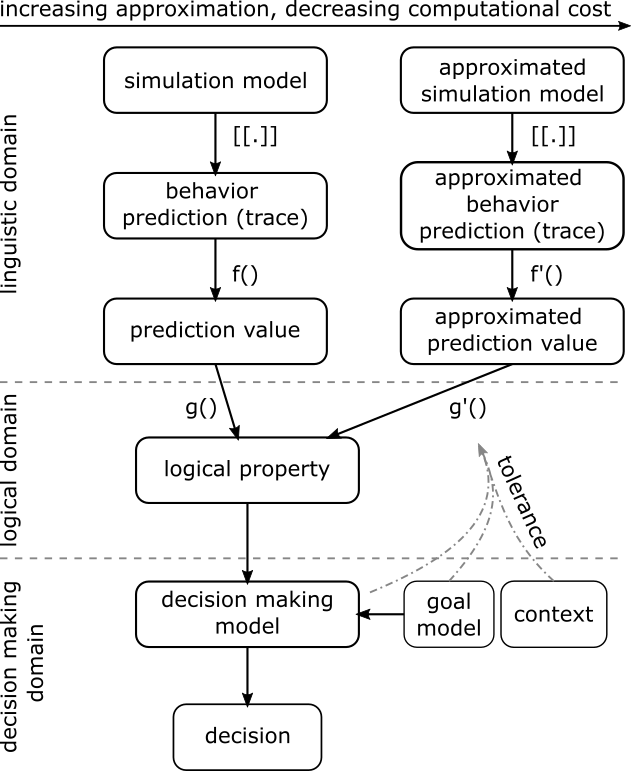}
    \caption{Conceptual Foundation for Model Approximation Technique from~\cite{biglari2022towards}.}
    \label{fig:conceptualfoundation over approximation}
\end{figure}

\begin{figure}
    \centering
    \includegraphics[width = 0.6\columnwidth]{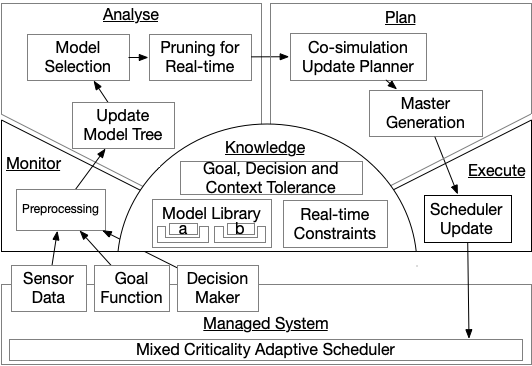}
    \caption{Real-Time Adaptive Abstraction and Approximation
Architecture from~\cite{biglari2022towards}.}
    \label{fig:MAPE_arch}
\end{figure}

However, for this adaptation between the different models, we need to establish that these models are substitutable in a specific situation. Figure \ref{fig:conceptualfoundation over approximation} represents the conceptual foundation from \cite{biglari2022towards} for reasoning about run-time adaptive approximation in a dynamic situation. This conceptual foundation is based on \cite{barroca2014integrating} that combines the use of ontology and language engineering.\\
A simulation of a model([[.]]) produces a trace. By analyzing this trace, we can use a function $(f())$ to calculate the system's prediction value related to a logical property. This prediction value is used in the decision-making process to make a decision. To reason about an allowed approximation of a model, we must consider the goal models, decision-making processes, and context. The sensitivities of the logical property in relation to decision-making in a specific context and for a specific goal must be mapped to enable this reasoning. The same reasoning process is applied when a different, more approximate decision-making model is used in a specific context.\\
We explore the impact of approximating a model on cost reduction as a foundation for a framework that enables real-time system adaptation. This framework allows for swapping models with more approximate versions based on the available library of models.
Utilizing the insights from the conceptual framework, Biglari et al. proposed an architecture for adaptive approximation in real-time systems. This architecture in figure\ref{fig:MAPE_arch} is based on MAPE-K architecture. MAPE-K is a high-level control loop for self-adaptive systems from IBM~\cite{1160055}.


An approximated model is most probably not valid in all the different contexts the system will be in. Consequently, the question arises: How can we do the adaptive approximation and abstraction at run-time, and find the most appropriate model for a specific context?\\
We use the Validity Frame concept to enable the adaptive approximation at run-time.

\subsection{Validity Frames}
The concept of frames has been present for some time, dating back to the early 1980s when the original idea of "experimental frames" was introduced by \cite{zeigler1984structures}.\\
\cite{klikovits2017modeling} addressed the structure of modeling frames, which depends on the activity performed, and this work describes why different activities require different frames. Extending the work of \cite{klikovits2017modeling},  \cite{van2021validity} formalizes the Validity Frame concept to specify the range-of-validity of a model. 

A Validity Frame ($VF$) contains all necessary data and processes for determining the appropriate usage of the model. A $VF$ is defined as a structure: \\
\begin{equation}
     VF =< S_M, \pi_M, \gamma_M, map_\pi\rightarrow S_M, map_\gamma\rightarrow	S_M, SPEC_{exe}, Val_M, \alpha_{val} >
\end{equation}

 Where:\\

$S_M$ is the model structure, which defines the set of model components and their relationship,
$\pi_M$ is the modeled properties.\\
$\gamma_M$ is the influencing factor that is
captured within $VF$ and enables reasoning about the utilization of the model in terms of what properties the model can provide valid answers for and under which influence factors.
Properties of the environment related to the system under study that can potentially influence a model's behavior and/or the range of validity concerning a particular property, even though they are outside of the model's control, are referred to as "influencing factors" \cite{mierlo2020exploring}. \\

$map_\pi$→$S_M$\ and $map_\gamma$→$S_M$ are explicitly mapped onto the model’s implementation.
The $SPEC_{exe}$ shows the specification of the simulation environment, the simulator, or the
specification of the embedded platform ($SPEC_{MBD}$) on which M is executed.
Model validity, $Val_M$, is a crucial part of $VF$, allowing reason about how well model M reflects the real-world counterpart $Sys$.

Last is the set of validation activities $\alpha_{val}$ includes a set of validation conditions $v$, used to validate or invalidate the behavior of M in known or unknown contexts.

A $VF$ is used to explicitly identify  the influence factors and  range-of-validity for a model and to provide methods and processes to ensure the model accurately represents the source system. This includes methods for calibrating the model, experiment design for validation, validation metrics, etc.\\
The purpose of validity frames is to clearly express the contexts in which a model generates valid results for a specific set of properties concerning a real-world system.\\

To convey different contexts, specific properties of the environment must be considered within our system.

\section{ADAPTIVE ABSTRACTION AND APPROXIMATION USING VALIDITY FRAMES} \label{sec: report}
In this section, we present the process we followed to implement the adaptive approximation technique using $VF$ in our case study.

\subsection{Case study: Lane Changing}

\begin{figure}[h!]
    \centering
    \includegraphics[width = 0.9\columnwidth]{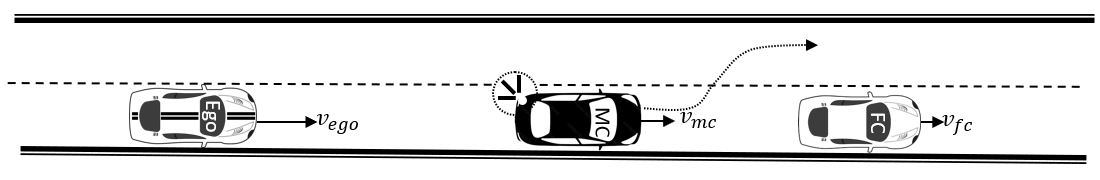}
    \caption{Scenario.}
    \label{fig:blinkingscenario}
\end{figure}

We use the Highway Lane change system as a case study to demonstrate our idea. A lane-changing system controls the movement of the ego (target car) in both the longitudinal (forward/backward) and lateral (left/right) directions. According to the environmental properties, the ego car changes or follows the lane.\\
This research presents a scenario of lane changing depicted in Figure \ref{fig:blinkingscenario}. We refer to this illustration throughout the paper to demonstrate our concepts and findings. In this example, there are three cars, ego car (Ego), middle car (MC), and front car (FC), which have the velocity of $v_{ego}$, $v_{mc}$, and $v_{fc}$ respectively. Moreover, $v_{fc}$ $<$  $v_{mc}$ $<$  $v_{ego}$ \\

To predict the positions of the different cars, the simulation can use several models with different levels of abstraction and approximation.
The first uses the kinematic equation $x(t) = 1/2*a*t^2 + v*t + x_0 $. In this formula, $a$ is the vehicle's acceleration in $m/s^2$, $v$, the velocity of the vehicle in $m/s$, and $x$, the vehicle's position. This model is the original model.
However, there is a possibility of using a more simplified model. We obtain this model by removing the acceleration term. We call this model the approximated model, which is less detailed but has a lower computational cost. These two models only take into account that the cars do not change lanes during the prediction of their position. 

In~\cite{biglari2022model}, the adaptive approximation is applied to a similar lane changing case study using these two different models. In the study of Biglari and Denil, a controller needs to select the most appropriate model at run-time between an original and approximated model. However, both models reason on the same set of properties and have the same environmental context, namely, the cars do not change lanes. 
Hence, selecting one valid model within the specific context is essential during model selection. The concept of a Validity Frame is highly beneficial in this regard.

The scenario in this paper describes an additional feature,  a turning or blinking light. A blinking light indicates the driver's intention to change lanes. In our running example, blinking is the influencing factor of interest, an environmental factor that is outside of the model's control but impacts model selection. 


We also added a much more computationally expensive model. This model simulates the lane changing behaviour of the different cars as well not only the ego car, and as such is valid both in the blinking and non-blinking light context. In this model, we simulate every car and considered each of them as an ego car, and run the simulation to perceive the behaviour of the system.

\subsection{Adaptation using a Validity Frame}
Our validity frame defines the different contexts of the system based on \emph{influencing factors} in which the original model and approximated models are valid.  For applying the adaptation with frames, two challenges arise:
\begin{description}
    \item[Challenge 1 - Properties of the environment and frames:]  Influencing factors are those properties of the environment that influence a model’s behavior and are quantified. We must identify these properties as factors that exert a significant impact. Once the properties are properly identified, characterised, a validity frame needs to be created. We will not focus on the aspect of building a frame but assume that proper techniques are available for creating the frame. 
    \item[Challenge 2 -  Run-time selection of models:] Having different models available, both abstracted, approximated and combinations, requires model management techniques. Organising these models is required. Furthermore, selecting the correct model at run-time requires proper data-structures and algorithms. 
\end{description}



\subsection{Design-time Model Organisation}

As a starting point, we use the Validity Frame Graph (VFG) proposed by Van Acker in ~\cite{van2021validity}. The validity graph encodes the relations between different models based on their validity. It is therefore classified as a mega-model~\cite{favre2006megamodelling}. 

Mega-modeling, as defined in \cite{favre2006megamodelling}, is the practice of creating a model that illustrates the global relationships between various modeling artifacts without focusing on their specific content. This is precisely what a Validity Frame Graph (VFG) does, capturing the abstract relationships between interrelated Validity Frames and their containing models without considering their specific modeling details.

However, in our case, the data-structure is quite naive, in a sense that all models are represented as a Vertex within the graph. Edges between the vertices encode how properties are removed, added, or the range of properties is changed between the models. The organisation in this data-structure results in a fully connected graph.

\begin{figure}[htb]
    \centering
    \includegraphics[width = 0.6\columnwidth]{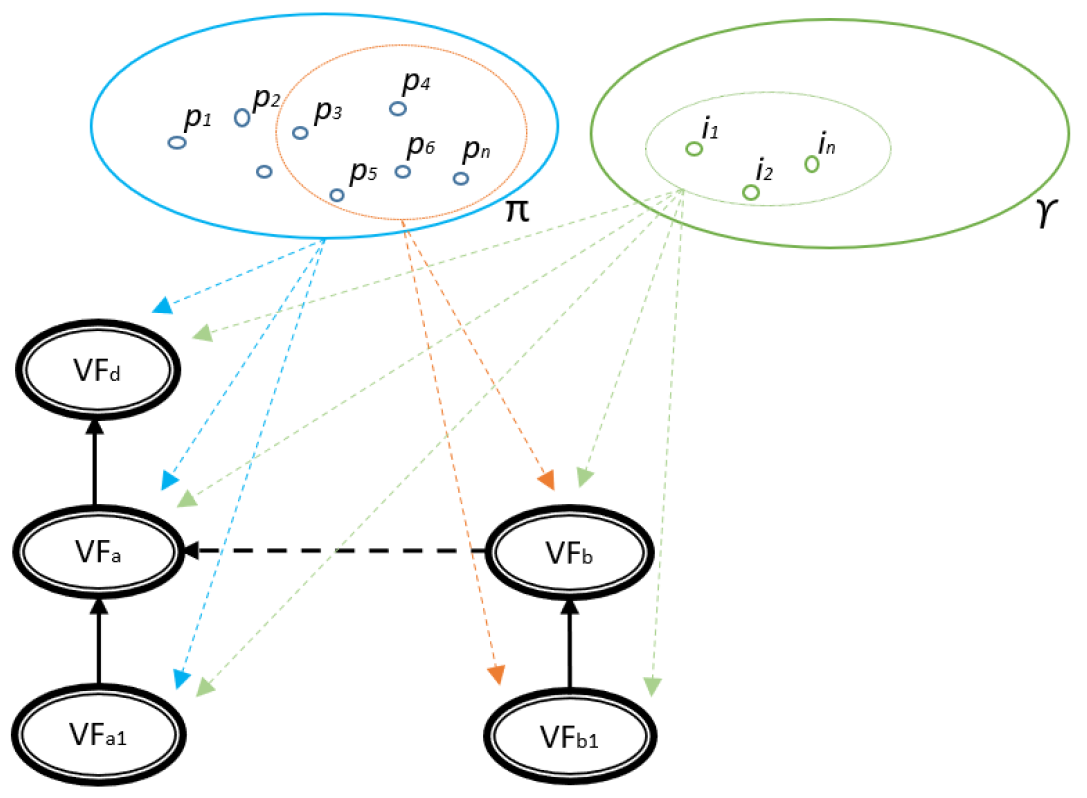}
    \caption{Validity Frame Graph example from \cite{bert2023journal}.}
    \label{fig:VFG_example}
\end{figure}

Figure \ref{fig:VFG_example} depicts an example of $VFG$. This figure shows five interrelated Validity Frames and corresponding models, which create a Validity Frame Graph. It visualizes the relationships between various models using the sets of properties and factors of influence, $\pi$, and $\gamma$. In the $VFG$, instances that depict the same $\pi$ and $\gamma$ are connected by an abstraction relation represented by a solid arrow. If an instance represents only a portion of the $\pi$ and/or $\gamma$, it is connected to the corresponding instance via a view decomposition relation represented by a dashed arrow \cite{bert2023journal}.

\begin{figure}[h!]
    \centering
    \includegraphics[width = 0.8\columnwidth]{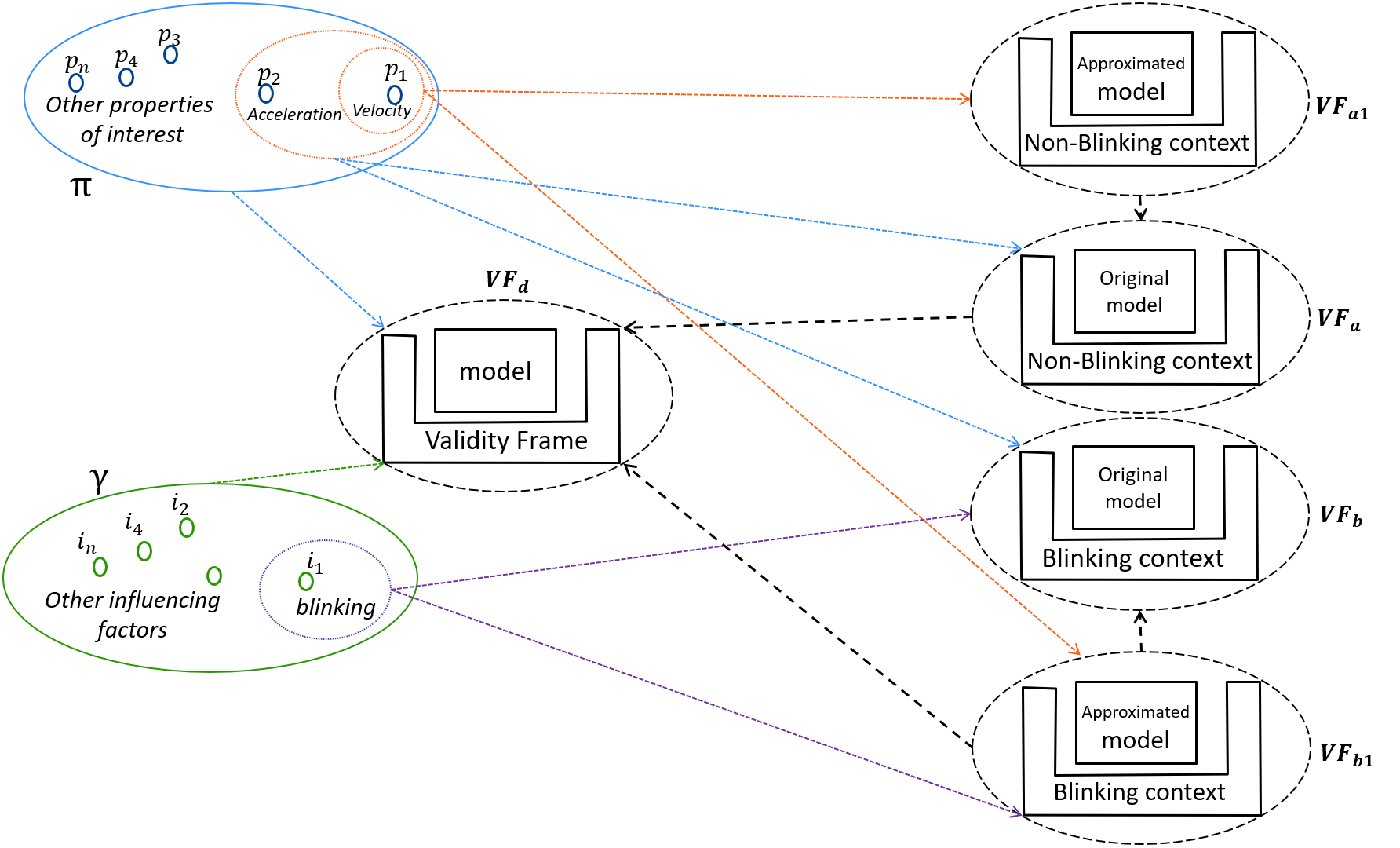}
    \caption{validity frames Graph for the Highway Lane change scenario.}
    \label{fig:vf_contex}
\end{figure}

Using the $VFG$ technique, we define and
classify models within a Validity Frame.
Figure \ref{fig:vf_contex} illustrates the $VFG$ for this Highway Lane change scenario. In Figure \ref{fig:vf_contex}, $\pi$ is the set of properties and $\gamma$ is the set of influencing factors. $VF_d$ is the head of the $VFG$, and the contained model, $M_d$ which is the most detailed model. The following relations are between
the contained models of the VFs:\\
\begin{itemize}
    \item[--] $M_a$ = $Abstract(M_d)$
    \item[--] $M_{a_1}$ = $Approximate(M_a)$
    \item[--] $M_b$ = $Abstract(M_d)$
    \item[--] $M_{b_1}$ = $Approximate(M_b)$
\end{itemize}
$M_a$ is the abstract version of $M_d$ and $M_{a1}$ is the approximated model based on $M_a$. And also $M_b$ is the abstracted model of $M_d$, and $M_{b1}$ is the approximated model based on $M_b$. 

\subsection{Run-time Model Organisation: Decision Tree}

\begin{figure}[htb]
    \centering
    \includegraphics[width = 1.05\columnwidth]{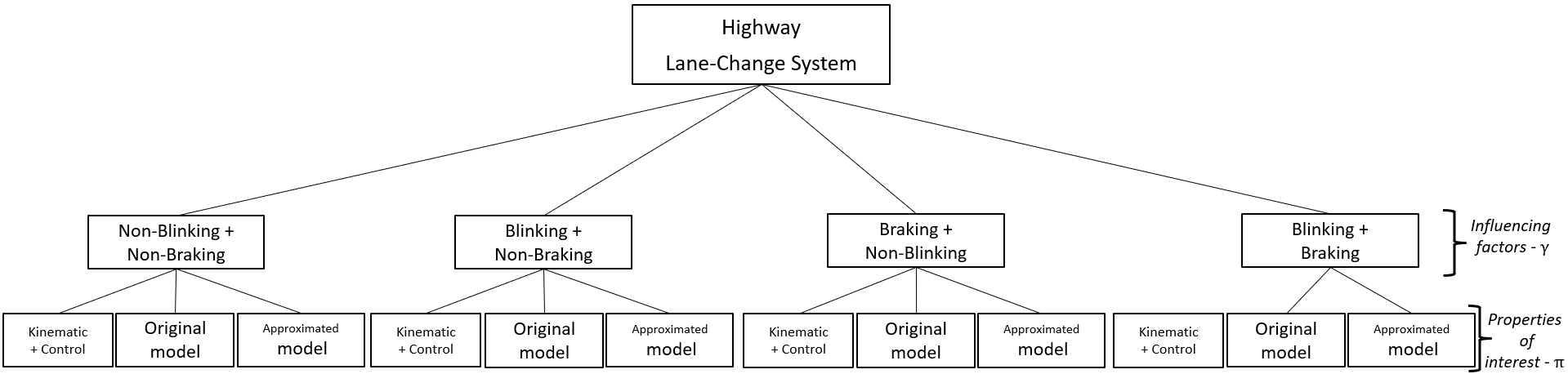}
    \caption{Decision Tree of Highway Lane Change Scenario.}
    \label{fig:decisionTree}
\end{figure}

As hinted in the previous section, a fully connected graph is not ideal for selecting models at run-time.
As such, another data-structure is needed. We utilize a decision tree to select the most suitable models for our system in a specific context, as depicted in figure \ref{fig:decisionTree}.

The organisation in a decision tree also requires a significant amount of domain knowledge. The designer needs to select in what order the properties and influence factors are ordered in the tree or possibly combined within a single splitting of the tree. 

To select a set of models, we use Decision Tree, and its time complexity is $O(depth)$. Using a fully connected graph needs choosing between Breadth First Search (BFS) and Depth First Search (DFS). The time complexity of the DFS and BFS graphs are represented in the form of $O(|V|+|E|)$, where $V$ is the number of nodes and $E$ is the number of edges. Therefore decision tree is computationally less expensive than BFS or DFS \cite{cormen2022introduction}.

We select the model during \emph{run-time} but build the decision tree at \emph{design time} to avoid increasing run-time computational cost. In Figure \ref{fig:decisionTree}, we traverse from the root to the leaf nodes according to the context. Afterward, we employ the response surface graph for the model selection method suggested by Biglari and Denil. The approximated model is a model with uncertainty and there is no possibility of using as much uncertainty as you like. So, you can switch to the approximated model if the uncertainty is within the bound. In \cite{biglari2022model}, the response surface graph shows how to find the bound and where you are allowed to swap between models. Algorithm \ref{alg} shows the pseudocode of selecting from the decision tree's leaf nodes and deciding to use the original or approximated model. This algorithm compares nextTrajectories, which correspond to different decisions. For example, the ego car may decide to change lanes or not, and if so, when and where. If simulations conclude the same decision, then we use the approximated model, which has less computational cost.\\

\begin{algorithm}
\caption{} \label{alg}
\begin{algorithmic}

\State $out1 \gets simulation\ result\ of\ the\ original\ model$ 
\State $out2 \gets simulation\ result\ of\ the\ approximated\ model$
   \\
    \If{$out1.nextTrajectories = out2.nextTrajectories$}
        \State $Use\ Approximated\ model.$
    \Else
        \State $Use\; Original\; model.$
    \EndIf
    
    
\end{algorithmic}
\end{algorithm}

Finally, to ensure that the model tree is optimized for real-time performance, we follow the MAPE-K control loop in figure \ref{fig:MAPE_arch}, and constraints are applied to eliminate infeasible solutions. As this is a real-time system, not all possible solutions are suitable. Thus, solutions that cannot meet deadlines are pruned.




\subsection{Experiments and Results of the Adaptation}

\begin{figure}[h!]
    \centering
    \includegraphics[width = 0.8\columnwidth]{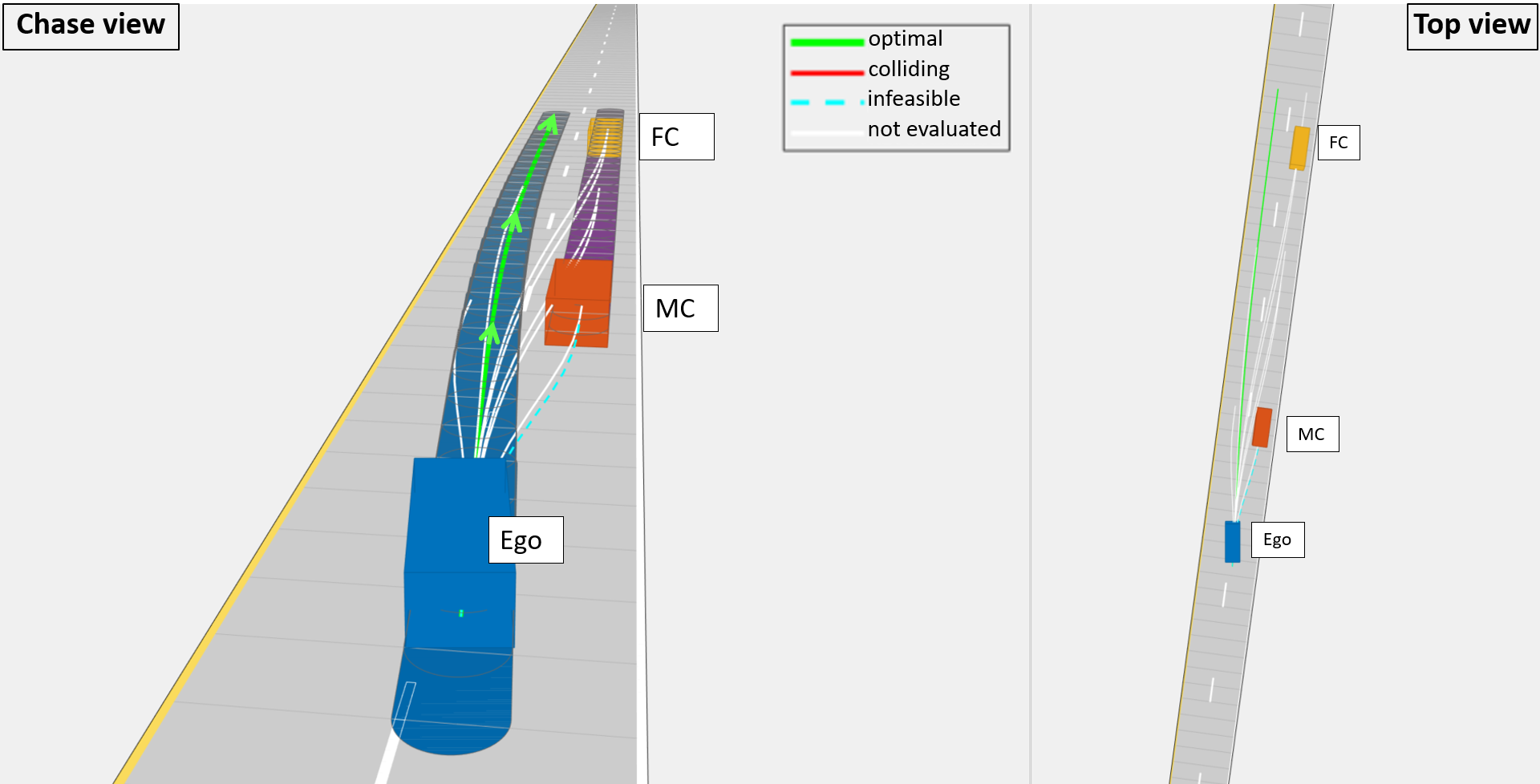}
    \caption{Non-Blinking Context Simulation.}
    \label{fig:resultnonblink}
\end{figure}

\begin{figure}[h!]
    \centering
    \includegraphics[width = 0.8\columnwidth]{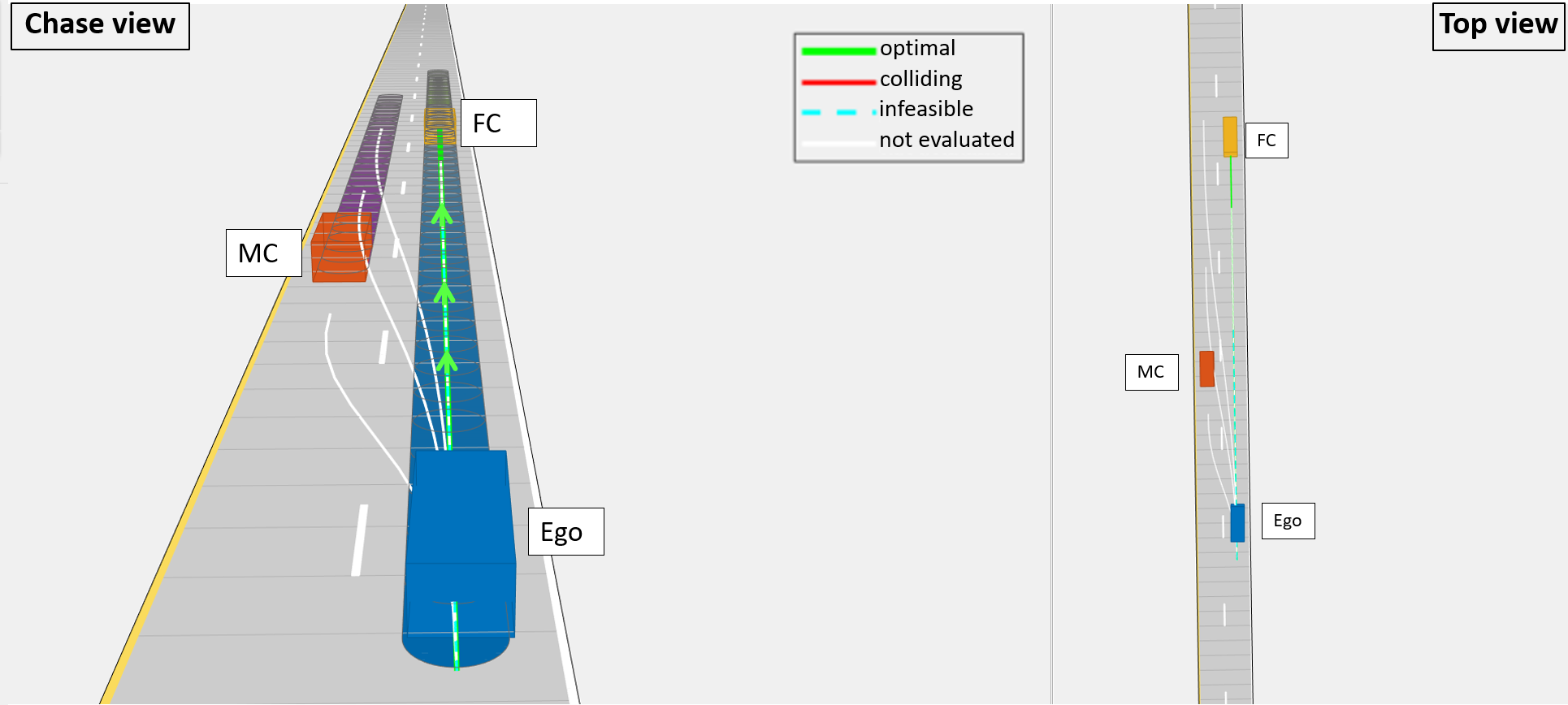}
    \caption{Blinking Context Simulation.}
    \label{fig:resultblink}
\end{figure}

In our running example scenario in Figure \ref{fig:blinkingscenario}, the velocity of $v_{ego}$, $v_{mc}$, and $v_{fc}$ are as follows: $v_{fc}$ < $v_{mc}$ < $v_{ego}$.
We show the simulation results for different contexts.
Figure \ref{fig:resultnonblink} shows the simulation result when there is a non-blinking context.
Accordingly, we add the influencing factor of blinking. Figure \ref{fig:resultblink} shows the simulation result in the blinking context. The optimal trajectory that the ego car follows is depicted by the arrowed line, while the other lines represent alternative trajectories that are deemed inappropriate.

The simulation results depict that in a blinking context when the ego car knows that MC intends to change the lane, in Figure \ref{fig:resultblink}, the ego car keeps the lane but in a non-blinking context, the simulation result is different, Figure \ref{fig:resultnonblink}, the ego car decides to change the lane.\\
Additionally, our scenario involves both an original and an approximated model. The simulation will determine if either model leads to a differing conclusion.
By using Algorithm \ref{alg} and response surface graph, if there is no different conclusion and the next trajectories are the same, the approximated model is considered suitable, as it has lower computational cost. Logically, using an approximated model instead of the original model decreases the computational cost because these models have a simpler structure with fewer parameters compared to more detailed models.


\section{DISCUSSION}  \label{sec:discussion}
In section \ref{sec: report}, we described the experiments and simulations we performed to reduce the computational cost while using the run-time adaptive approximation technique. Consequently, we chose the more appropriate model from our library of models and interrelated Validity Frames for our system under study.

In this section, we discuss some of these steps, presenting lessons learned and highlighting additional questions raised and identified challenges:

\begin{description}

    \item[Challenge A: A library of related models' data-structure.] We use the library of related models. Accordingly, we need to find out what such a library of related models looks like. In our case, we used the validity frame graph at design time and the decision tree at run-time.  Multiple models are created for a single system, $Sys$. By clearly defining the validity of each model within its Validity Frame ($VF$), using a Validity Frame Graph $VFG$ \cite{bert2023journal} can visualize the relationships between the models.
    To overcome challenge A, we also need to solve some sub-challenges.\\

    \begin{description}
    
        \item[Encoding relations between properties.] There are three types of properties in a system: \cite{bert2023journal} system properties, properties of interest, and Influencing factors (environment properties).
            \begin{itemize}
                \item System properties: related to the real-world entity. It is part of the possibly infinite set of properties required to grasp all possible contexts and facets of the considered real-world entity. In practice, one cannot fully specify this property set, as it is impossible to capture all object’s facets in all its possible contexts.
                \item Properties of interest. The set of properties of interest  gives rise to the abstraction relation between model M and system Sys: a model only provides (correct) answers with respect to these properties, while other properties are abstracted away. More specifically, the implemented set of properties of model M is a subset of the system properties.
                \item Influencing factors. The set of influence factors can also influence the abstraction relation  between model M and system Sys: a model is only supposed to provide answers concerning the modeled properties under the modeled influence factors, while other influence factors are abstracted away. More specifically, the implemented set of influence factors of the model is a subset of the system properties.
        \end{itemize}

        Therefore, properties of interest and influencing factors are properties that we consider. After bringing out properties in our example are distance and blinking, the next step is finding the relation between these properties. How to encode this relation is still the question.

        \item[Order in the property set.]

        There seems to be an intuitive concept on ordering in the set of models that exist. However, the ordering can only be partial as we cannot really compare different properties and their ranges to each other and make a conclusion if the model has a bigger validity. As such a partially ordered set or poset might be a better visualisation. In a poset, each element can be compared to each other element and results in a bigger, smaller, equal or incomparable result. However, when we create relations per property, we get a fully ordered relation. Although these options still need more thinking, handling multiple properties is still one of the unsolved challenges.

    \end{description}
    \item[Challenge B: How to select a model at run-time to use in a specific situation?]  We can not use all our models at run-time in all different situations (context). Then two questions pop up.

    \begin{description}
        \item[Search Algorithm.] We use Decision Tree so the time complexity is $O(depth)$. The designer must decide on the arrangement of properties and impact factors in the tree or combine them into a single splitting. There are some splitting decision tree methods to choose from.

        \item[Translation beforehand to a data-structure] CPS is a real-time system, so we can not do everything at run-time and tend to increase the computational cost at run-time. To avoid the overhead at run-time, we propose to do translation beforehand to a data-structure at design time that is more feasible. For instance, making a decision tree at design time.

        \end{description}

    \item[Challenge C: Can we automatically convert a mega-model to a decision tree (or related data-structure)?] Currently, converting a mega-model to a decision tree is performed manually as we fully understand our system, but this may not be the case in other environments and systems. 
    As previously mentioned, the order of the decisions and organisation of properties and influence factors within the decision tree affects its run-time performance. The challenge is to automatically create a decision tree that is optimal to use at run-time. Figure \ref{fig:decisionTree} represents our first idea of implementing the decision tree. In this data structure, level 1 of the tree includes influencing factors. The nodes would be sorted down as the number of branches in level 1 is less than level 2 of the tree, which is depicted in Figure \ref{fig:DecisionTreeIdea}.

    \begin{figure}[htb]
    \centering
    \includegraphics[width = 0.9\columnwidth]{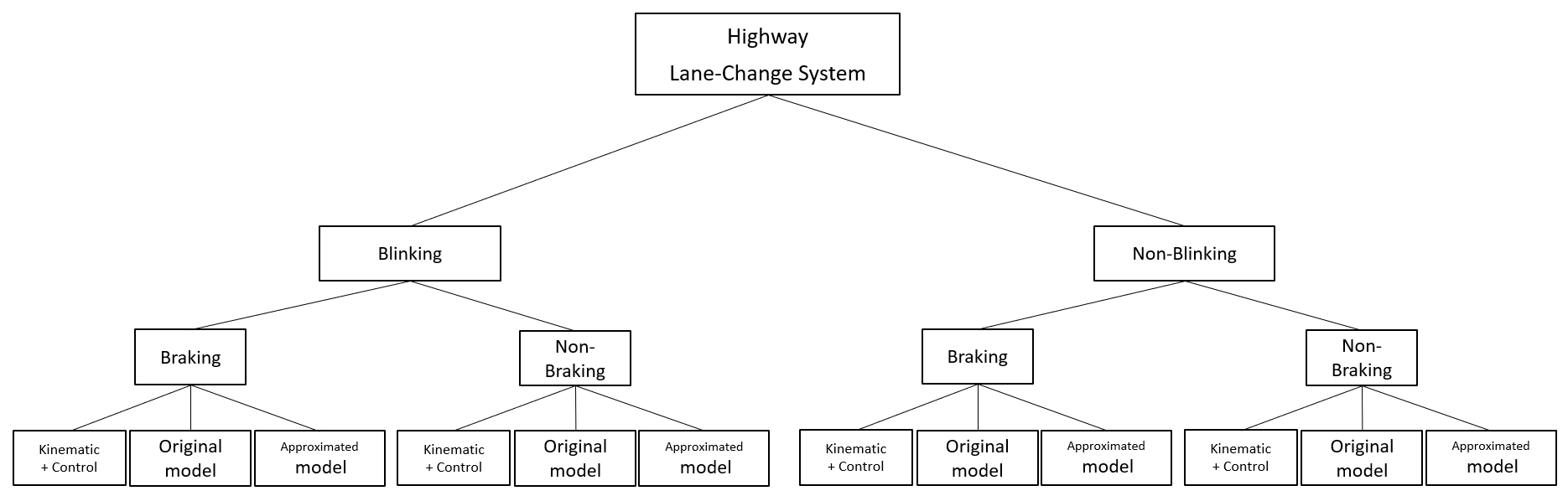}
    \caption{Decision Tree Representation.}
    \label{fig:DecisionTreeIdea}
\end{figure} 

    For building the decision tree, we utilize properties of interest and influencing factors, and there are relationships between these properties. However, there is a need to optimise the organisation of the tree. For this, there is a need for domain knowledge about which contexts are accessed often.   A \textbf{\emph{heat map}} is a graphical representation of data to visualize the relationship between variables.
    Heat maps might help discover the feature importance of each property in the tree. 
    A heat map of feature importance provides a clear view of which properties have the greatest impact and which play a less significant role. Figure \ref{fig:heatmap} represents the heatmap where the possibility of non-braking/non-blinking context is more than other ones.\\
    
        \begin{figure}[htb]
    \centering    \includegraphics[width = 0.5\columnwidth]{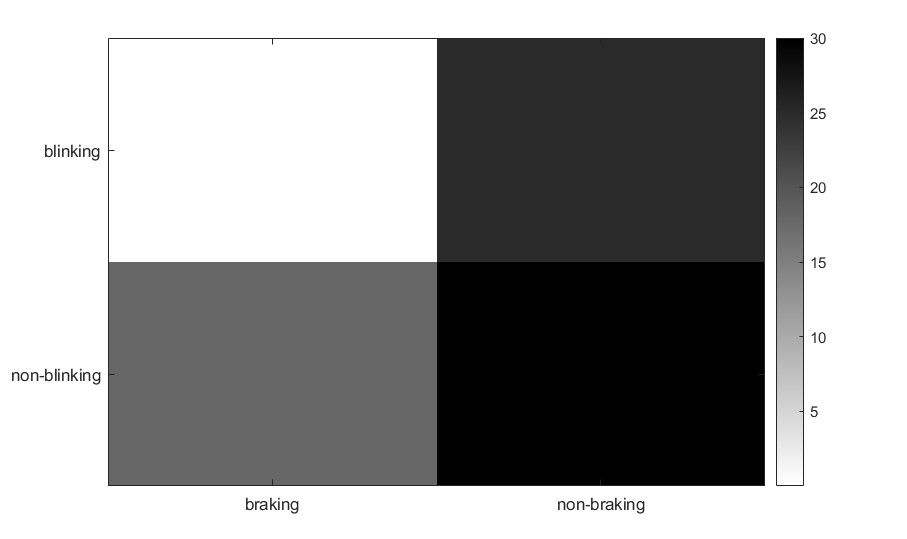}
    \caption{Heat Map.}
    \label{fig:heatmap}
\end{figure} 

    Furthermore, it is essential to show that the resulting tree can be used in all different contexts as the mega-model. However, this does not mean that all models in the mega-model should be present in the run-time data structure. This is closely related to the next challenge.

    \item[Challenge D: Which surrogate models to create?] It is possible to create a set of surrogate models starting from a single model. Typically, the smaller the context, the less computational effort is required to simulate the model. However, creating a surrogate model, either by hand or computationally, requires a tremendous amount of effort. Balancing the number of models and  their operating range will be needed. As such, guidelines and heuristics will need to be created to assist designers in making appropriate decisions. Again, the domain knowledge about often used contexts is important to see which surrogates to create for the model at design time. Information from the operation phase of the system might be needed to make optimised decisions. Digital twins are needed for such a setup. 

\end{description}

This experience report acknowledges that validation is a crucial step in modeling. However, for this study, we proceed with the assumption that the models have been thoroughly validated and are functioning as intended.

In our running example, we used blinking as the influencing factor of interest to represent the validity frame concept. There are other influencing factors that we could add to our validity frame, for example, traffic, road and weather conditions.

\section{CONCLUSIONS} \label{sec:conclusion}
Real-time adaptive approximation in Cyber-Physical Systems is a method used to reduce the computational cost of running models. However, an approximated model is not valid in all the various contexts the system may operate in.
In this experience report, we applied the adaptive approximation technique using the Validity Frames concept in our case study.
We utilized the concept of Validity Frames to enable adaptation at run-time in different contexts. 
In this work, we used the properties of interest and environment properties known as influencing factors within a validity frame, and the relations between different validity frames to choose the most appropriate model from our library of models.
We discussed identified challenges, lessons learned, and additional questions raised.\\



\section*{Acknowledgments}
Raheleh Biglari is funded by the BOF fund at the University of Antwerp.

\bibliographystyle{apalike}
\bibliography{bibliography}

\section*{Author Biographies}

\textbf{\uppercase{RAHELEH BIGLARI}} is a doctoral student at the University of Antwerp, Faculty of Applied Engineering in Electronics and ICT. Her email address is \textit{raheleh.biglari@uantwerpen.be}.

\textbf{\uppercase{Joachim Denil}} is an associate professor at the University of Antwerp, Faculty of Applied Engineering in Electronics and ICT. His email address is \textit{joachim.denil@uantwerpen.be}.

\end{document}